\documentclass[aps,floatfix,footinbib, twocolumn,superscriptaddress,amsmath,amssymb,amsfonts]{revtex4-1}

\usepackage[T1]{fontenc}
\usepackage{lmodern} 
\usepackage{color}
\usepackage{bbold}
\usepackage{dsfont}
\usepackage{graphicx}
\usepackage[caption=false]{subfig}
\usepackage[colorlinks]{hyperref}
\usepackage[bottom]{footmisc}
\usepackage{enumerate}
\usepackage{float}

\usepackage{empheq}
\usepackage{tikz}
\usepackage{fancyhdr}
\usepackage[normalem]{ulem}
\DeclareMathAlphabet\mbc{OMS}{cmsy}{b}{n}
\usepackage{balance}

\begin{document}

\global\long\def\eqn#1{\begin{align}#1\end{align}}
\global\long\def\vec#1{\overrightarrow{#1}}
\global\long\def\ket#1{\left|#1\right\rangle }
\global\long\def\bra#1{\left\langle #1\right|}
\global\long\def\bkt#1{\left(#1\right)}
\global\long\def\sbkt#1{\left[#1\right]}
\global\long\def\cbkt#1{\left\{#1\right\}}
\global\long\def\abs#1{\left\vert#1\right\vert}
\global\long\def\cev#1{\overleftarrow{#1}}
\global\long\def\der#1#2{\frac{{d}#1}{{d}#2}}
\global\long\def\pard#1#2{\frac{{\partial}#1}{{\partial}#2}}
\global\long\def\re{\mathrm{Re}}
\global\long\def\im{\mathrm{Im}}
\global\long\def\dd{\mathrm{d}}
\global\long\def\ddd{\mathcal{D}}

\global\long\def\avg#1{\left\langle #1 \right\rangle}
\global\long\def\mr#1{\mathrm{#1}}
\global\long\def\mb#1{{\mathbf #1}}
\global\long\def\mc#1{\mathcal{#1}}
\global\long\def\tr{\mathrm{Tr}}
\global\long\def\dbar#1{\Bar{\Bar{#1}}}

\global\long\def\nth{$n^{\mathrm{th}}$\,}
\global\long\def\mth{$m^{\mathrm{th}}$\,}
\global\long\def\non{\nonumber}

\newcommand{\orange}[1]{{\color{orange} {#1}}}
\newcommand{\cyan}[1]{{\color{cyan} {#1}}}
\newcommand{\blue}[1]{{\color{blue} {#1}}}
\newcommand{\yellow}[1]{{\color{yellow} {#1}}}
\newcommand{\green}[1]{{\color{green} {#1}}}
\newcommand{\red}[1]{{\color{red} {#1}}}
\global\long\def\todo#1{\orange{{$\bigstar$ \cyan{\bf\sc #1}}$\bigstar$} }

\title{Emergence of Non-Gaussian Coherent States Through Nonlinear Interactions}

\author{M. Uria}
\affiliation{Departamento de F\'{\i}sica,  Facultad
de Ciencias F\'isicas y Matem\'aticas, Universidad de Chile,
Santiago, Chile}
\affiliation{ANID - Millenium Science Iniciative Program - Millenium Institute for Research in Optics}
\author{A. Maldonado-Trapp}
\affiliation{Departamento de F\'isica, Facultad de Ciencias F\'isicas y Matem\'aticas, Universidad de Concepci\'on, Concepci\'on, Chile}
\author{C. Hermann-Avigliano}
\affiliation{Departamento de F\'{\i}sica,  Facultad
de Ciencias F\'isicas y Matem\'aticas, Universidad de Chile,
Santiago, Chile}
\affiliation{ANID - Millenium Science Iniciative Program - Millenium Institute for Research in Optics}

\author{P. Solano}
\affiliation{Departamento de F\'isica, Facultad de Ciencias F\'isicas y Matem\'aticas, Universidad de Concepci\'on, Concepci\'on, Chile}

\begin{abstract}
Light-matter interactions that are nonlinear with respect to the photon number reveal the true quantum nature of coherent states. We characterize how coherent states depart from Gaussian by the emergence of negative values in their Wigner function during the evolution while maintaining their characteristic Poissonian photon statistics. Such states have non-minimum uncertainty yet present a metrological advantage that can reach the Heisenberg limit. Non-Gaussianity of light arises as a general property of nonlinear interactions, which only requires a polarizable media, resonant or dispersive. Our results highlight how useful quantum features can be extracted from the seemingly most classical states of light, a relevant phenomenon for quantum optics applications.
\end{abstract}

\maketitle

{\it Introduction.---}Oscillating charges, antennas, and lasers naturally radiate electromagnetic (EM) waves in coherent states \cite{Glauber1963,Glauber1963b}. These states have a well-defined phase and amplitude, and, for intense sources, they are represented by nearly a point in the quadrature (phase) space, giving the deceiving impression of being classical states of light. However, their quantumness is buried in their phase and amplitude noise, i.e., their quantum fluctuations. Such inherent, albeit hidden, quantum nature can be unveiled upon interaction with matter \cite{Hacker2019,Uria2020}, leading to squeezed or non-Gaussian phase space representations, more evident signatures of quantum features. Historically, states with such features are broadly dubbed non-classical.

Non-classical states of light can be harvested for a range of uses, from increasing the precision of a measurement to processing information beyond what is possible with classical resources ~\cite{Arrazola_2021, Casacio_2021,Anderson17,Giovannetti_2011,HarocheBook}.
However, such quantum states are naturally uncommon, and generating them requires a variety of specialized experimental settings that rely on nonlinear interactions between the EM field and matter \cite{Strekalov2019,yan2021,Villas-Boas2020,Prasad2020}. 
Considering the applications and demand for intense, macroscopic, non-classical light across different platforms, it becomes relevant to understand the minimal conditions to extract useful quantum properties from the more common, yet still quantum, coherent states of light.

This paper describes light-matter interactions via a model Hamiltonian that is nonlinear in the photon number operator. We define a family of states of the EM field that fully characterizes the evolution of a macroscopic coherent state under such interactions. The state of the field becomes highly quantum and  non-Gaussian upon evolution while maintaining its Poissonian photon statistics and average intensity. We analyze the non-Gaussianity and metrological advantage of such states through the negativity of the Wigner function~\cite{Kenfack2004,Siyouri2016} and the quantum Fisher information~\cite{Braunstein94}, respectively. The results show that any nonlinear interaction universally leads to highly quantum non-Gaussian coherent states of light, regardless of the specifics of the matter subsystem. Finally, we discuss the implications and outlook of our results before the concluding remarks.

{\it Model.---}
The light-matter interaction Hamiltonian in the dipole approximation is given by $\hat{H}_{\rm{int}}=-\mathbf{\hat{P}}_{\rm{a}}\cdot\mathbf{\hat{E}}$,
where $\mathbf{\hat{E}}$ is the electric field and $\mathbf{\hat{P}}_{\rm{a}}$ is the total atomic dipole moment. Depending on the atomic susceptibility $\chi$, the interactions are generally nonlinear in the electric field, as $\hat{P}_{\rm{a}}\propto\sum_i\chi^{(i)}\hat{E}^{i}$. For an initial coherent state of the field in the large intensity limit, atoms and field evolve almost as separable states \cite{Gea-Banacloche1991,Chumakov1994,Chumakov1995,Saavedra1998}. In such case, the interaction Hamiltonian in the interaction picture and rotating wave approximation can be approximated by a product of the atomic operator and a field operator proportional to the square-root of the photon number, $\hat{E}\sim\sqrt{\hat{n}}$, as $\hat{H}_{int}=\hbar g_i \hat{S}_x\hat{n}^{\frac{i+1}{2}}$, where $g_i$ includes the information of the $i$-th order susceptibility, $\hat{S}_{(x,y,z)}$ is the collective atomic spin operator, and we have assumed the atomic frequency to be $i+1$ times the field frequency \footnote{This particular condition is equivalent to a degenerate parametric amplification of an arbitrary order, which could be selected by imposing the resonant condition of the field in a cavity.}. To study a range of nonlinear light-matter interactions for single-mode macroscopic states of the EM field, independently of the details of the matter subsystem, we consider the following interaction Hamiltonian in the interaction picture
\begin{equation}
    \hat{H}_{\rm{int}}=\hbar g \hat{n}^{\epsilon}\hat{O}_a,
    \label{eq:H}
\end{equation}
where $\hbar$ is the reduced Planck constant, $g$ is the coupling frequency, $\hat{O}_a$ is an atomic operator, and $\epsilon$ is the nonlinear parameter , determined by the type of interaction.

Although Eq. (\ref{eq:H}) excludes some nonlinear interactions, such as the squeezing Hamiltonian, it represents the simplest form of nonlinearities and provides a useful representation of a range of physical interactions. For example: the Jaynes-Cummings Hamiltonian in the dispersive limit, when $\epsilon=1$ and $\hat{O}_a=\hat{S}_z$\cite{HarocheBook}; the Kerr Hamiltonian, when $\epsilon=2$ and $\hat{O}_a$ is a non-zero constant number~\cite{NonClassicalStates}; and the Jaynes-Cummings Hamiltonian in the resonant intense field limit, when $\epsilon=1/2$ and $\hat{O}_a=\hat{S}_x$ \cite{Uria2020,Gea-Banacloche1991,Chumakov1994,Chumakov1995,Saavedra1998}. Particularly, $\epsilon=(i+1)/2$ is related to a resonant $i$-th order susceptibility when $\hat{O}_a=\hat{S}_x$, and its definition helps to show that non-classical light emerges for nonlinearities in the photon number operator.

We assume that the initial state of the light-matter system $\ket{\psi(0)}$ is the outer product of the field in a coherent state $\ket{\alpha}$ and an arbitrary atomic pure state $\sum_j c_j \ket{\lambda_j}$, where $\ket{\lambda_j}$ are the eigenstates of $\hat{O}_a$. The Schr{\"o}dinger equation for the evolution of $\ket{\psi(t)}$  leads to the separable state \footnote{The overall approximation of the interaction Hamiltonian in Eq. (\ref{eq:H}) and the separability of the state at any time $t$ are valid for as long as $g t<\bar{n}$ \cite{Gea-Banacloche1991,Chumakov1994}, equivalent to the intense field limit we are studying.}
\begin{equation}
\ket{\psi(t)}=\sum_j c_j \ket{\alpha_{\theta_j,\epsilon}(t)}\otimes \ket{\lambda_j},
\label{eq:a+f_state}
\end{equation}
where
\begin{equation}
    \ket{\alpha_{\theta,\epsilon}(t)}=\sum_n \frac{\alpha^n e^{-|\alpha|^2/2}}{\sqrt{n!}} \rm{Exp}\left\{-i \theta n^{\epsilon}t\right\}\ket{n},
    \label{GCS}
\end{equation}
and $\theta_j=g\lambda_j$ is a function of the light-matter coupling strength and the eigenvalues of the atomic operator.

The states $\ket{\alpha_{\theta,\epsilon}(t)}$ constitute a subset of the previously called generalized coherent states (GCS) \cite{Titulaer1965b,Bialynicka1968,Stoler1971,Yurke1986}, and in this work we refer to them as such. As we will show next, they represent the evolution of the EM field from a Gaussian (standard) to a non-Gaussian coherent state, maintaining the photon statistics while offering a metrological quantum advantage.

\begin{figure}[t]
 \centering
  \includegraphics[width=0.48\textwidth]{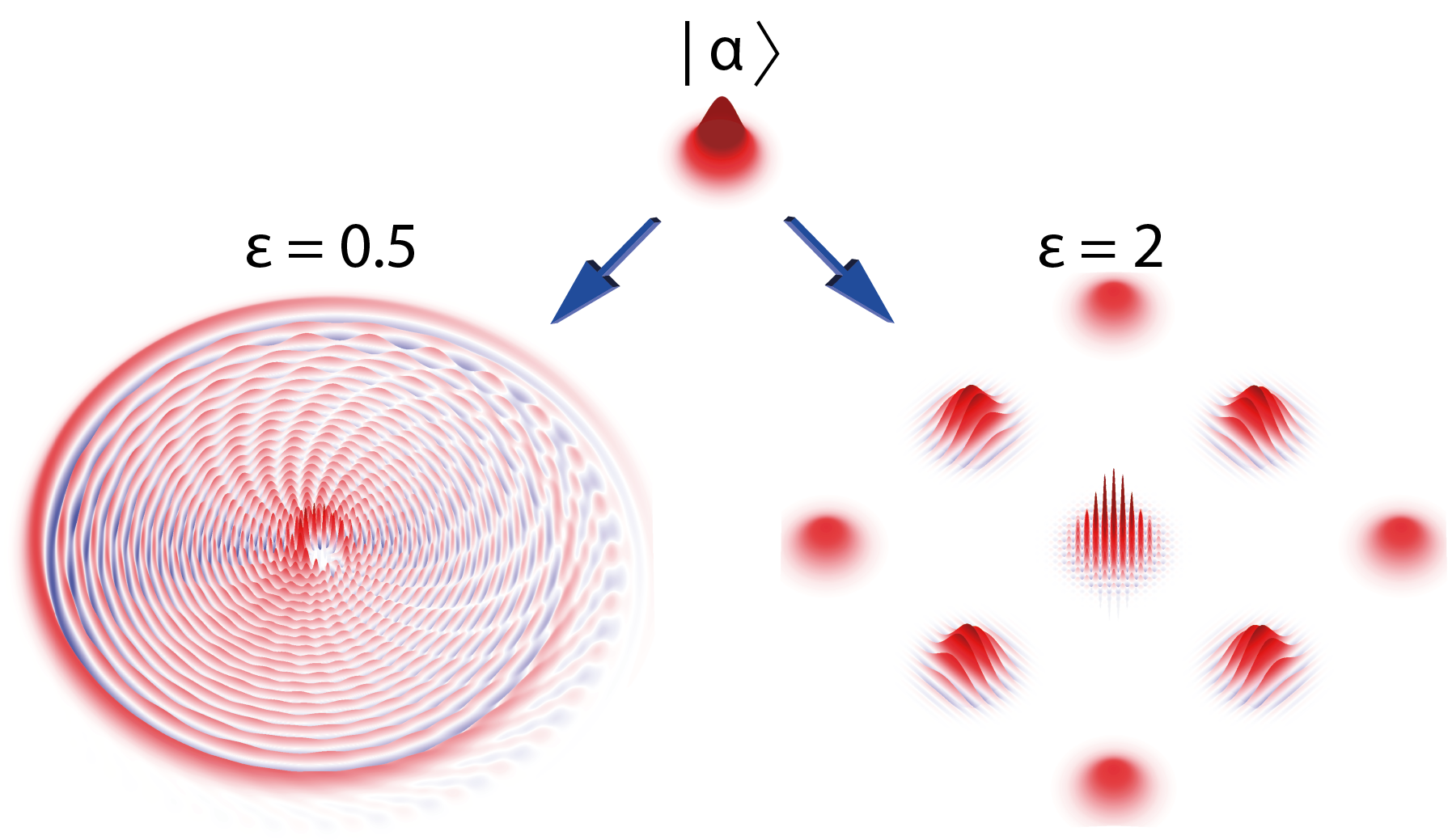}
 \caption{Examples of possible evolutions of an intense, macroscopic, coherent state of the electromagnetic field upon nonlinear interaction with matter. The Wigner functions represent the state of the field $\ket{\alpha}$, with $|\alpha|^2=50$, that evolves from a coherent to a highly quantum non-Gaussian coherent state, depending on the nonlinear parameter  $\epsilon$. The color red (blue) represents positive (negative) values of the Wigner function.}
 \label{fig:schematics}
\end{figure}

{\it Generalized coherent states.---}
The GCS represent a family of states that fully characterizes the evolution of a coherent state under nonlinear light-matter interactions of the form of Eq. (\ref{eq:H}). They reduce to well-known states depending on the details of the interaction. For example: $\ket{\alpha_{0,\epsilon }(t)}=\ket{\alpha}$, $\ket{\alpha_{\theta,0}(t)}=e^{-i \theta t}\ket{\alpha}$, $\ket{\alpha_{\theta,1}(t)}= \ket{\alpha e^{-i \theta t}}$, and $\ket{\alpha_{\theta,2}(t)}$ represents Kerr states \cite{NonClassicalStates}, which have been largely studied for their potential applications in quantum metrology. Fig. \ref{fig:schematics} exemplifies two possible evolutions of the Wigner function of GCS for $\epsilon=0.5$ and $2$. The figures share the same scale, which exemplifies how a minimum-uncertainty, nearly localized, state can spread over the quadrature space creating a, evidently quantum, non-Gaussian field.

GCS are non-orthogonal and they form an overcomplete basis, with $\int d\alpha \ket{\alpha_{\theta,\epsilon}(t)}\bra{\alpha_{\theta,\epsilon}(t)}=\pi\mathbb{1}$, where $\mathbb{1}$ is the identity operator. Their mean photon number is constant throughout the evolution, $\bra{\alpha_{\theta,\epsilon}(t)}\hat{n}\ket{\alpha_{\theta,\epsilon}(t)}=|\alpha|^2$. In general, the expectation value of any power of the photon number operator respect to the GCS is the same as the one respect to a coherent state, meaning $\bra{\alpha_{\theta,\epsilon}(t)}\hat{n}^{m}\ket{\alpha_{\theta,\epsilon}(t)}=\bra{\alpha}\hat{n}^{m}\ket{\alpha}$\cite{Agarwal1992}. In particular, their Mandel $Q$ parameter \cite{Mandel79} and its generalization to higher order statistics \cite{Lee90,Kim02} corresponds to the same photon statistics of a coherent state. More properly defined, given the $n$-order correlation function of the field $G^{(n)}$, a state is said to have coherence of the $m$-order if $|G^{(n)}|^2=\Pi^{2n}_{j=1}G^{(1)}$ for $n \leq m$ \cite{Glauber1963,Titulaer1965a}, equivalent to say that the $n$-th order normalized correlation function is equal one. Such condition is satisfied to all orders for the GCS, earning them the name coherent. Furthermore, GCS do not show minimum-uncertainty or squeezing in any quadrature \cite{Stoler1971,Agarwal1992}. This means that any non-Gaussianity or quantum advantage that they might present does not come from removing or adding a quanta of light to the field or changing its Poissonian photon statistics. It rather comes from the modification of the quantum fluctuations of the electric field.

{\it Non-Gaussianity of generalized coherent states.---}
We characterize the non-Gaussianity of the EM field by the condition of a negative value in its Wigner function \cite{Kenfack2004,Siyouri2016}, which indicates quantum features. The Wigner negativity is defined as $\mathcal{N}_\phi=\int d\beta\left[ |W^{\ket{\phi}\bra{\phi}}(\beta)|-W^{\ket{\phi}\bra{\phi}}(\beta)\right]$, where $W^{\ket{\phi}\bra{\phi}}$ is the Wigner function of $\ket{\phi}$.

The Wigner functions of the GCS, given by Eq. (\ref{GCS}), are:
\begin{multline}
    W^{\ket{\alpha_{\theta,\epsilon}}\bra{\alpha_{\theta,\epsilon}}}(\beta) = \frac{2 e^{-\alpha^2-2r^2}}{\pi}\left( e^{-\alpha^2+4\alpha r \cos{\delta}}-\right.\\
    4\sum_{m,n,m>n}(-1)^n \frac{\alpha^{m+n}}{m!}    \sin{\left((m-n)\delta +
    \frac{\theta}{2} (m^\epsilon-n^\epsilon)\right)}\\
    \left.\sin{\left(\frac{\theta}{2} (m^\epsilon-n^\epsilon)\right)}  (2r)^{m-n} L_n^{m-n}(4r^2) \right),
    \label{eq:W}
\end{multline}
where $L_n^{m}$ are the generalized Laguerre polynomial and $\beta=re^{i\delta}$ is a complex number that represents every point in quadrature space.

\begin{figure}[t]
\centering
 \includegraphics[width=0.8\linewidth,keepaspectratio]{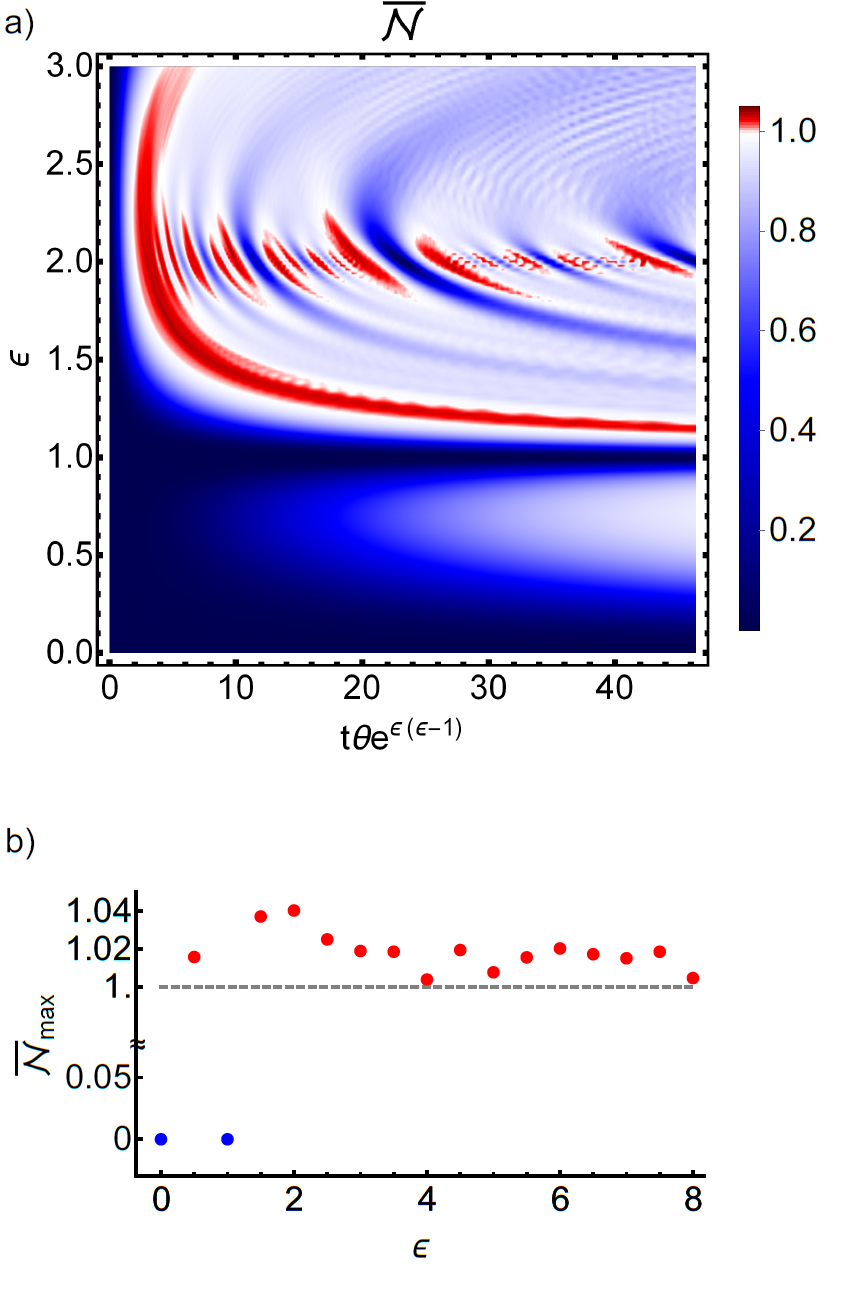}
\caption{a) Wigner negativity $\bar{\mathcal{N}}$ as a function of the evolution parameter $\theta$ for different nonlinear parameters $\epsilon$ considering an initial coherent state with an average photon number of $\bar{n}=10$. The evolution parameter is re-scaled by $e^{\epsilon(\epsilon-1)}$ for visualization purposes. b) Maximum Wigner negativity as a function of the nonlinear parameter. The red (blue) dots represent non-zero (zero) negativity. $\bar{\mathcal{N}}$ is normalized to that of a Fock state with $n=10$ for both figures.}
\label{fig:N}
\end{figure}

Figure \ref{fig:N} a) shows the evolution of the Wigner negativity $\mathcal{N}$ of GCS for different values of the nonlinear parameter  $\epsilon$, numerically calculated from Eq. (\ref{eq:W}). We normalize the negativity by that of a Fock state with the same mean photon number. We observe that as long as there is a degree of nonlinearity ($\epsilon\notin\{0,1\}$), the field will eventually become non-classical \footnote{We point out that the time at which the Wigner negativity for $\epsilon<1$ is maximum is not shown in the figure for visualisation purposes.}. Fig. \ref{fig:N} b) shows the maximum Wigner negativity reached through the entire evolution for different values of $\epsilon$. For all nonlinear interactions, the negativity of the GCS can surpass that of a Fock state. This result opens a discussion about the interpretation of the negativity of the Wigner function and how, or if, quantumness could be quantified, which is beyond the scope of this paper.

{\it Quantum advantage of coherent states.---}
The metrological advantage of the GCS can be quantified by their quantum Fisher information $\mathcal{F}^{\rm{Q}}$ \cite{Braunstein94}.  It relates to the minimum attainable uncertainty in estimating a parameter $\delta x$ via the Cramer-Rao bound $\sqrt{\langle\delta x^2\rangle}\geq 1/\sqrt{\bar{n}\mathcal{F}^{\rm{Q}}}$, where $\bar{n}$ is the mean number of photons involved in the measurement.
When the quantum Fisher information scales with the average number of photons, one achieves the minimum uncertainty $\sqrt{\langle\delta x^2\rangle}\propto 1/\bar{n}$, which is the optimum scenario for quantum metrology known as the Heisenberg limit \cite{Tan2019,Escher2011}. GCS are good candidates to detect small displacements of the field \cite{Penasa2016,Lewis-Swan2020}. 
The quantum Fisher information for displacements in the phase space of a pure state is given by the variance of the quadrature in the direction orthogonal to the displacement, namely $\mathcal{F}^{\rm{Q}}=4\left<(\Delta X)^2\right>$ \cite{Escher2011}. 

The quadrature variance of a GCS is
\begin{equation}
    \left<(\Delta X)^2\right>_{\alpha_{\theta,\epsilon}}(t)=4\bar{n}\left( \left< Re[z_{2}(t)]^2\right>_\alpha - \left< Re[z_{1}(t)]\right>_\alpha^2 \right)+1
\end{equation}
where we define the expectation value $\langle o_n\rangle_\alpha=\sum_n |\langle\alpha|n\rangle|^2 o_n$ and the function $z_{j}(t)=\rm{Exp}\{-i\theta(t)((n+j)^\epsilon-n^\epsilon)\}$.
GCS evolve between states of large and small Fisher information, with $1\leq\left<(\Delta X)^2\right>_{\alpha_{\theta,\epsilon}}\leq4\bar{n}+1$. The $\bar{n}$ dependence in the variance, and consequently $\mathcal{F}^{\rm{Q}}$, leads to the Heisenberg limit. We highlight that the maximum Fisher information for GCS is equivalent to that of a squeezed state with a squeezing of $-10 \rm{Log}_{10}\left[4\bar{n}+1\right]$ dB \footnote{For $\bar{n}>8$ one can generate states with metrological advantage beyond what has been experimentally achieved with squeezed light \cite{Vahlbruch2016,Ulrik2016} }, while displaying non-minimum uncertainty and Poissonian photon statistics.

\begin{figure}[t]
\centering
 \includegraphics[width=0.8\linewidth,keepaspectratio]{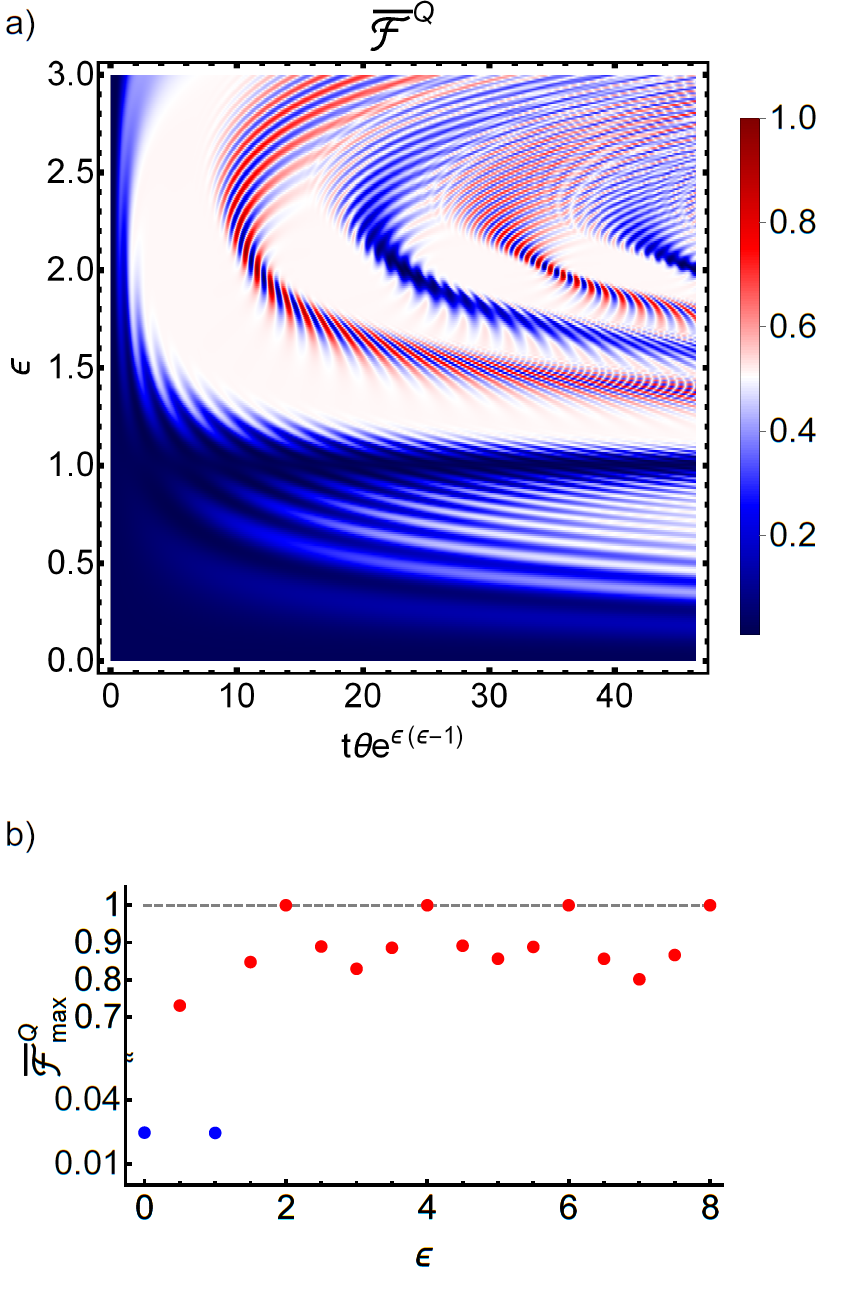}
\caption{a) Quantum Fisher information $\bar{\mathcal{F}}^{\rm{Q}}$ and as a function of the evolution parameter $\theta$ for different nonlinear parameters $\epsilon$ considering an initial coherent state with an average photon number of $\bar{n}=10$. The evolution parameter is re-scaled by $e^{\epsilon(\epsilon-1)}$ for visualization purposes. b) Maximum quantum Fisher information as a function of the nonlinear parameter. The red (blue) dots represent metrological advantage (or lack thereof). $\bar{\mathcal{F}}$ is normalized to its maximum value, $4(4\bar{n}+1)$. }
\label{fig:F}
\end{figure}

Figure \ref{fig:F} a) shows the evolution of the quantum Fisher information  $\bar{\mathcal{F}}^{\rm{Q}}$ for different nonlinear parameters, normalized by its highest value. The quantum Fisher information from the variance of the orthogonal quadrature is equivalent, but with the red-to-blue fringes in the figure completely out of phase. Fig. \ref{fig:F} b) shows the maximum quantum Fisher information reached through the entire evolution for different nonlinear parameters. We observe a metrological advantage over a coherent state for all nonlinear parameters, reaching an optimum $\bar{\mathcal{F}}^{\rm{Q}}_{\rm{max}}=1$.

Both Figs. \ref{fig:N} and \ref{fig:F} show the evolution of a coherent state with an average photon number of $\bar{n}=10$. Nonetheless, the overall behavior of $\bar{\mathcal{N}}$ and $\bar{\mathcal{F}}^{\rm{Q}}$ is independent of $\bar{n}$. For the range of numerically tested parameters, $\bar{\mathcal{N}}_{\rm{max}}>1$ for all nonlinear interactions. Similarly, the dynamic behavior of $\bar{\mathcal{F}}^{\rm{Q}}_{\rm{max}}$ always oscillate between 0 and 1 for even values of $\epsilon$ and offers metrological advantage over a coherent state for $\epsilon\neq\{0,1\}$. This result shows that non-classicality emerges for arbitrarily large, \textit{i.e.} macroscopic, initial coherent states of the EM field. 

{\it Role of the atomic subsystem.---}
So far, we have analyzed the case of the EM field interacting with a medium consisting of one or many atoms in a pure state that is the eigenstate of the collective atomic spin operator represented by $\hat{O}_a$ in Eq. (\ref{eq:H}). Such description leaves out the effects of the atomic state on the EM field. To address this issue, we consider that the atomic subsystem starts in an arbitrary mixed state $\rho_a(0)=\sum_{i,j} c_{i}c_{j}^*\ket{\lambda_i}\bra{\lambda_j}$. Then, the state of the EM field evolves to
$\rho_f(t)=\sum_j |c_{j}|^2 \ket{\alpha_{\theta_j,\epsilon}(t)}\bra{\alpha_{\theta_j,\epsilon}(t)}$, a statistical mixture of the GCS. Such mixed states have diminished quantum properties compared to pure GCS, but they can still be non-classical.

To quantify the role of the initial state of the atomic subsystem, we consider it to be a Werner state defined as \cite{Werner1989,wilde2017}
\begin{equation}
    \rho_{a}(0)= p\ket{\Psi}\bra{\Psi} + (1-p) \frac{1\otimes 1}{d},
\end{equation}
where $d$ is the dimension of the Hilbert space of the atomic subsystem, and $\ket{\Psi}=\sum_j c_j\ket{\lambda_j}$ is an arbitrary pure state written in terms of the eigenstates $\ket{\lambda_j}$ of the collective atomic operator $\hat{O}_a$. The parameter $p$ allows to continuously vary the initial state from fully mixed ($p=0$) to pure ($p=1$). The state of the EM field after its evolution is
\begin{equation}
    \rho_f(t)=\sum_j \left(p |c_j|^2 + \frac{1-p}{d}\right)\ket{\alpha_{\theta_j,\epsilon}(t)}\bra{\alpha_{\theta_j,\epsilon}(t)}.
    \label{eq:Werner}
\end{equation}
Since the Wigner function is a linear transformation, we get that
 \begin{equation}
     W^{\rho_f}(\beta,t)=\sum_j \left(p |c_j|^2 + \frac{1-p}{d}\right) W^{\ket{\alpha_{\theta_j,\epsilon}(t)}\bra{\alpha_{\theta_j,\epsilon}(t)}}(\beta).
     \label{eq:Wigner_Werner}
 \end{equation}
 
\begin{figure}[t]
 \centering
  \includegraphics[ width=0.45\textwidth]{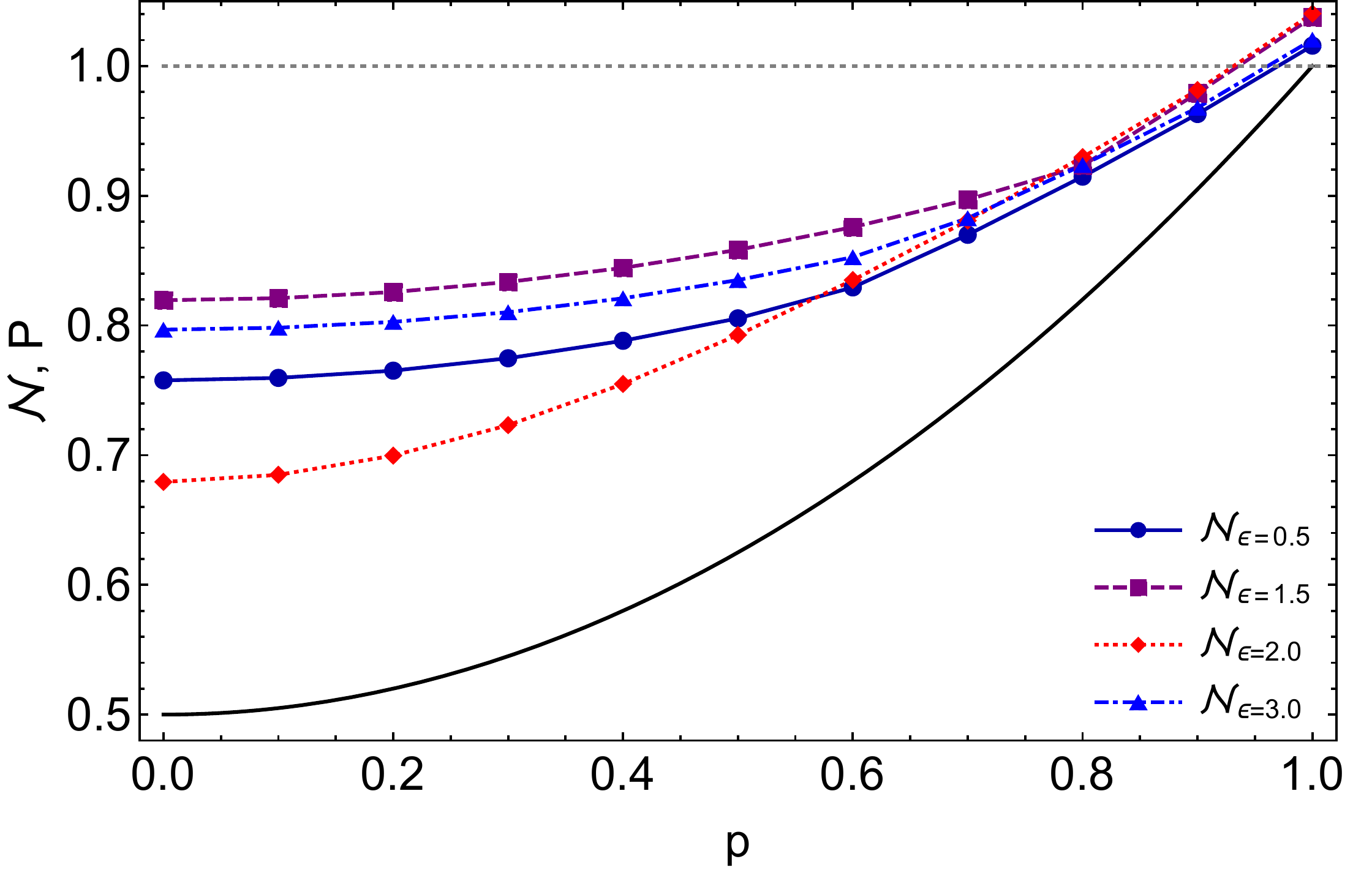}
 \caption{Maximum Wigner negativity $\bar{\mathcal{N}}_{\rm{max}}$ of the EM field and purity $P$ of the initial atomic state (solid black) as a function of the Werner parameter $p$. The negativity for the nonlinear parameters $\epsilon=\{0.5,1.5,2,3\}$ is shown with circles, triangles, diamonds, and squares, respectively. The Wigner negativity is normalized to that of a Fock state with the same average photon number.}
 \label{fg:WernerW}
\end{figure}

Figure \ref{fg:WernerW} shows the maximum normalized Wigner negativity $\bar{\mathcal{N}}_{\rm{max}}$ during the EM field evolution as a function of the Werner parameter $p$. The initial atomic state in the case of highest purity ($p=1$) corresponds to an eigenstate of the atomic operator, leading to a pure GCS state of the EM field. The Wigner negativity increases monotonically with $p$, same as the purity of the initial atomic state $P=Tr\{\rho_a(0)^2\}$. 

From Eq. (\ref{eq:Wigner_Werner}) one sees that for $|c_j|^2=1/d$ the Wigner function of the EM field does not depend on $p$. This means that the obtained state of the field is the same whether the initial atomic state is fully mixed or a pure state with an equal superposition of eigenstates of the atomic operator $\hat{O}_a$. Moreover, for $d=2$, the maximally entangled states are equal superpositions of the eigenstates of the collective atomic operators $\hat{S}_{(x,y,z)}$. This suggests that quantum correlations between the atoms are not just unnecessary but can be detrimental for obtaining a non-classical EM field.

We notice that even for $p=0$, the resulting mixed state of the EM field still presents a non-zero Wigner negativity for $\epsilon\neq\{0,1\}$, evidencing the ubiquity of non-classical light emerging from nonlinear interactions. We highlight the universal aspect of the evolution into non-classicality since the maximum negativity of the resultant state of the EM field is relatively insensitive to the nonlinear parameter $\epsilon$, especially towards high purity of the initial atomic state.

{\it Discussion.---}
Macroscopic coherent states of light are commonly considered the most classical ones. However, nonlinear interactions can be used to extract their quantum resources even in a closed system with no initial quantum correlations and unaffected by measurements \cite{Teklu2015,Albarelli2016}. Our results are another example of this phenomenon, which can be interpreted as the nonlinear evolution unevenly spreading the minimal uncertainties of a coherent state throughout the quadrature space, while keeping a phase coherence among the EM field components.


This phenomenon offers new possibilities for engineering the EM field. For example, different linear combinations of the GCS can lead to useful states. In particular, combining two GCS with $\epsilon=1/2$ one could obtain a macroscopic Fock state, as Ref. \cite{Uria2020} discusses.



{\it Conclusions.---}
We studied an effective Hamiltonian for light-matter interactions that are nonlinear in the photon number operator. When the electromagnetic field begins in an intense coherent state, it evolves into a macroscopic highly quantum state. We present a family of states that characterizes the field evolution for all degrees of nonlinearity. We described some of their properties: i) invariant Poissonian photon statistic, ii) non-Gaussianity, characterized by negative values of their Wigner function, and iii) metrological advantage, which can reach the Heisenberg limit. We finally show that initial correlations in the atomic subsystem are unimportant, and maximally mixed states also lead to non-Gaussian coherent states of light. Any degree of nonlinearity is enough to extract quantum features from a coherent state, regardless of the details of the interaction and the medium. In that sense, non-Gaussianity emerges from nonlinearity as a universal phenomenon. We believe that understating the details of this process can contribute towards improving the engineering of quantum states of the electromagnetic field for quantum metrology and quantum information science applications.

{\it Acknowledgments.---} 
We are grateful to Alberto Marino, Kanupriya Sinha, Pablo Barberis-Blostein and Pierre Meystre for insightful comments and discussions. This work was supported in part by FONDECYT grants N$^{\circ}$ 11190078 and N$^{\circ}$ 11200192, ANID-PAI grants 77190033 and 77180003, and ANID - Millenium Science Inititive Program - ICN17\_012.


\bibliography{ref}

\end{document}